\documentstyle[11pt] {article}
\renewcommand{\thefootnote}{\fnsymbol{footnote}}
\def\zid{1\kern-0.36em\llap~1}
\def\u{\uparrow}
\def\d{\downarrow}
\def\r{\rangle}
\def\su{SU$_q(2)$}
\def\l{\lambda}
\def\m{${\cal M}$ }

\newcommand{\beq}{\begin{equation}}
\newcommand{\ber}{\begin{eqnarray}}
\newcommand{\eeq}{\end{equation}}
\newcommand{\eer}{\end{eqnarray}}

\begin{document} \begin{titlepage}
\rightline{\vbox{\halign{&#\hfill\cr
&\normalsize ANL-HEP-PR-92-12\cr
&\normalsize March 1992 \cr
hep-th/9203027 \cr   }}}
\vspace{.5in}
\begin{center}

\LARGE {\bf Altering the Symmetry of Wavefunctions in Quantum Algebras and
Supersymmetry }
\vskip   .6in
\Large C.K. Zachos\\
\vskip .4in
\small
High Energy Physics Division\footnote{Work supported
by the U.S.Department of Energy, Division of
High Energy Physics, Contract W-31-109-ENG-38}
\\Argonne National Laboratory, Argonne, IL 60439-4815, USA\\
zachos@anlhep      \end{center}

\vskip .7in

\begin{abstract}
The statistics-altering operators \m present in the limit $q=-1$ of
multiparticle \su-invariant subspaces parallel the action of such operators
which naturally occur in supersymmetric theories. We illustrate this
heuristically by comparison to a toy  $N=2$  superymmetry  algebra,
and ask whether there is a supersymmetry structure underlying \su ~at
that limit.  We remark on the relevance of such alternating-symmetry
multiplets to the construction of invariant hamiltonians.
\end{abstract}
\renewcommand{\thefootnote}{\arabic{footnote}} \end{titlepage}

Even though the suitability of quantum algebras [1] to particle physics has not
yet been fully understood, there has been a spate of speculation in
non-co-commutative contexts in solid state, molecular,  and nuclear physics.

The new generalized permutation symmetries involved in quantum algebras [2]
have led to several connections [3,4] to anyons [5] and parastatistics.
It might even appear,  deceptively,  that q-deformations provide a smooth
interpolation between bosons and fermions.  But, in previous work [6] on
deformed multioscillator algebras, we have stressed the crucial
difference between mere anticommutativity of some wavefunctions
(easy to arrange through deforming maps of boson operators)
versus nilpotence (the Pauli exclusion principle), which remains the
only genuine hallmark of fermionic behavior, as elaborated
more than two generations ago by Jordan, Wigner, L\"uders, and Klein.
As a result, q-deformations do {\em not} automatically blend graded and
conventional bosonic algebras into  the same structure, and each has to be
deformed separately; the noncommutative plane and quantum superplane are,
of course, distinguishable.

Nevertheless, by studying the ${\cal R}$-matrix which codifies the symmetries
of composite states, Saleur [7] has observed intriguing connections of
conventional \su-invariant models to supersymmetry, for generic parameter $q$.
Specifically, he discovered a ``hidden supersymmetry" Osp$_q(1|2)$ in, e.g.,
vertex models, leading to a quasi-graded classification of their states:
\su-symmetric vertex models map to Osp$_{-q^2}(1|2)$ graded models for
representations of the same dimension.

Here, in a different vein, we focus on statistics-altering operators \m in
q-algebras in the particular extreme case $q=-1$, and briefly note their formal
similarity with operators of this type that arise naturally in extended
supersymmetric theories, illustrated with a toy $N=2$ supersymmetry algebra.
Specifically, within the same \su ~multiplet, we observe raising/lowering
operators changing the symmetry of the wavefunction in a manner analogous
to the action of pairs of different supersymmetry charges. This phenomenon is
expected to also occur in more general contexts. However, if there is an
underlying supersymmetry in that limit, $q=-1$, it has not been  isolated.

Recall Drinfeld--Jimbo's standard \su ~quantum deformation:
\beq
[J_0,J_+]=J_+  \qquad \qquad [J_+,J_-]={1\over 2}~ [2J_0]_q
\qquad \qquad [J_-,J_0]=J_-~,
\eeq
where   $[x]_q \equiv ( q^x -q^{-x})/ (q-q^{-1}) $, the
``$q$-deformation'' of $x$.  The Casimir invariant of this algebra is
\beq
C_q \equiv 2J_+ J_- +[J_0]_q [J_0 -1 ]_q =2J_- J_+ +[J_0]_q [J_0 +1 ]_q ~.
\eeq
Recall that the classical limit  $q \rightarrow 1$ yields SU(2).  Further note
that (also self-evident from [8]), for generic $q$, the irreducible
representations of dimension $2j+1$ for $j$ integer or half-integer
yield the eigenvalue $[j]_q [j + 1]_q$  for this Casimir invariant.

The relevant feature of this deformed algebra is that the standard [1]
coproduct algebra homomorphism
\beq
\Delta (J_0)=J_0 \otimes \zid +\zid \otimes J_0 \qquad \qquad \qquad
\Delta (J_{\pm})=J_{\pm} \otimes q^{J_0} + q^{-J_0}\otimes J_{\pm}~,
\eeq
is non-cocommutative, in contrast to that of the classical limit
(addition of angular momenta), i.e.~it differentiates between
leading and trailing representation vector spaces.
Since the coproduct does not commute with the permutation group,
the representations of the deformed algebra will not be expected to be
characterized by conventional permutation symmetries. (cf. [2] for
q-modifications of permutations).

Apply the above coproduct to combine two spin-1/2 doublets,
${\bf 2}\otimes{\bf 2} ={\bf 3} \oplus {\bf 1}$:
The singlet is
\beq
\alpha= |q^{1/2} \u \d - q^{-1/2} \d\u\r,
\eeq
while the triplet is
\ber
\beta = |\u\u\r  \\
\Delta(J_{-}) \beta = {1\over\sqrt{2}}~|q^{1/2} \u \d + q^{-1/2} \d\u\r  \\
\Delta(J_{-})^2 \beta = |\d\d\r.
\eer
The normalization of $\alpha$ and $\Delta(J_{-})\beta$ is $1/\sqrt{[2]_q}$
for real $q^2$, but we ignore inessential normalizations for simplicity, on
account of the complex cyclotomic limits to  follow presently.

Now consider the $q=-1$ limit.  Polychronakos [3] has noted that
this case, among $p$-th roots of unity, provides a singular exception to the
dimensionality restriction of representations [9], essentially
because $[p/2]_q$ is nonvanishing for $p=2$. In fact, the q-algebra itself
reduces to classical SU(2), the coproduct remaining non-cocommutative
all the same. (The raising/lowering generators are hermitean conjugates,
$J_{+}^{\dagger}=J_{-}$,  for half-integral spin representations, whence
unitarity; but antihermitean conjugates---``negative parity" in the
terminology of ref.[10]---for integral spins, comprising an SU(1,1);
renormalizing to hermitean ones in this case reverts to SU(2) unitary
representations.)  For integral spins, the above Casimir eigenvalue is
negative, $-j(j+1)$, but for half-integral spins, it diverges to
$ 4/(q-1/q)^2 +1/2 + j(j+1)$.
The discussion to follow may thus be thought of as the exploration of this
unconventional composition law for SU(2)-representations.

The above states become:
\ber
\alpha= |i\u \d +i\d\u\r,
\eer
and
\ber
\beta = |\u\u\r  \\
\Delta(J_{-}) \beta = {1\over\sqrt{2}}~ |i\u \d -i\d\u\r  \\
\Delta(J_{-})^2 \beta = |\d\d\r.
\eer
By $\Delta(J_{+}) |\d\d\r = -\Delta(J_{-})\beta $ and
$\Delta(J_{+})^2 |\d\d\r  = \beta $, it is evident in this case
that  $\Delta(J_{+})^{\dagger}= - \Delta(J_{-})$, so that
renormalizing $\Delta(J_{+})  \rightarrow -\Delta(J_{+})$ leads to the
conventional spin-1 representation of SU(2).\footnote{Predictably,
the universal-${\cal R}$, $U$, and Clebsch matrices of e.g. ref.[11] turn
out different than those of the $q=1$ limit.}
In general, the unitary representation of SU(2) is provided by
$(-)^{2j+1} J_{+},~ J_{-}, ~J_0$.

Now note that the above singlet wavefunction $\alpha$ is symmetric, and the
members of the triplet are symmetric, antisymmetric, symmetric,
respectively. The symmetry of the wavefunction alternates {\em within} a
multiplet, but the dimensionality of multiplets has not changed. The
raising and lowering operators in the coproduct act as statistics-altering
operators \m. It is crucial for this alternation of symmetry that the
spins entering into the coproduct be half-integral.

Have the constituents of the states been converted to fermions? No, as
is apparent in the multi-isodoublet wavefunctions below.
Nevertheless, this alteration of the symmetry of the wavefunction is
reminiscent of supersymmetry. To illustrate this, introduce the elementary
supersymmetry algebra consisting of the (graded) direct product of two
``supersymmetric quantum mechanics" algebras:
\ber
s s^{\dagger} +s^{\dagger} s=\zid \\
S S^{\dagger} + S^{\dagger} S=\zid\\
S^{\dagger} S^{\dagger} =s^{\dagger} s^{\dagger} = SS=ss=0\\
sS+Ss=0\qquad\qquad s^{\dagger} S^{\dagger}+ S^{\dagger} s^{\dagger} =0  \\
sS^{\dagger}+ S^{\dagger} s=0\qquad \qquad
s^{\dagger} S+Ss^{\dagger} =0.
\eer

(The reader may recall that this familiar nonsimple graded Lie algebra is a
Wigner-Inon\"u contraction of the simple one SU($2|1$) [12].
In terms of Gell-Mann's standard SU(3)-basis matrices, SU($2|1$) consists
of four fermionic generators ($\l_4,\l_5,\l_6,\l_7$) and four
boson generators $(\l_1,\l_2,\l_3,\l_8')$,
where $\l_8'=\zid-\sqrt{3}/4 \l_8=$diag(3/4,3/4,3/2), so that it is
supertraceless like the rest. Define $\l_4\pm i\l_5$ to be $s^{\dagger},s$
and $\l_6\pm i \l_7$ to be $S^{\dagger},S$.
To contract, scale $\l_3,\l_8'$ by an expandible factor $r$;
$\l_1,\l_2$ by $r^{3/4}$; and all fermionic generators by $r^{1/2}$.
As $r\rightarrow \infty$, all commutators of the algebra trivialize, and so
do the cross-anticommutators between $s^{\dagger}(s)$ and $S(S^{\dagger})$;
whereas the anticommutators $\{s^{\dagger},s\}$ and $\{S^{\dagger},S\}$ are
unaltered. The now central generators $\l_3,\l_8'$ may be absorbed into the
normalization of the four fermion generators, resulting into the above
$N=2$ algebra.)

This supersymmetry algebra may be realized on two boson states $|B\r,~|b\r$,
and two fermion states $|F\r~,|f\r$, so that:
$S|B\r=|F\r ,~ s|b\r=|f\r,~S^{\dagger} |F\r=|B\r,~s^{\dagger}|f\r=|b\r$,
the remaining actions being null: $S|F\r= S|b\r =s|B\r=s^{\dagger}|F\r
=s^{\dagger}|b\r=S^{\dagger} |f\r=0 $, etc. Then:
\ber
s|Bb+bB\r =|Bf+fB\r   \\
Ss|Bb+bB\r=|Ff-fF\r.
\eer
So $\Delta (J_{-})$ switches the symmetry of the wavefunction like the even
(bosonic) operator $Ss=-sS$, but only the latter and not the former is
nilpotent. Morevover, there appear  no odd (fermionic) states or operators in
the former case (\su), although a less direct connection cannot be
excluded.  It may be tempting to ask whether some sort of underlying
supersymmetry could become manifest after appropriate fermionization.

Just as in the case of conventional SU(2), combining $n$ doublets for any $q$
yields the  Clebsch-Gordan decomposition series (the number preceeding the
boldface  representation dimensionality label indicates multiplicity of that
representation  in the reduction):
\beq
{\bf 2}^{\otimes n} = \bigoplus_{k=0}^{\lfloor n/2 \rfloor}~
\Bigl( {n+1-2k \over n+1}  {n+1 \choose k}\Bigr)~~({\bf n+1-2k})~,
\eeq
where $\lfloor n/2 \rfloor$ is the integer floor function. The coproduct
used is the iterated composition of Eqn.(3), unique by  co-associativity.

For instance, for spins {\bf 2}$\otimes{\bf 2}\otimes{\bf 2}={\bf 4}
\oplus{\bf 2}\oplus{\bf 2}$,  we have, unnormalized, a Quartet:
\ber
|\u\u\u\r  \\
           |q \d\u\u +\u\d\u +q^{-1} \u\u\d \r   \\
           |q^{-1} \u\d\d +\d\u\d +q\d\d\u \r\\
           |\d\d\d\r,
\eer
a Doublet I:
\ber
          |q^{-1/2} \d\u\u - (q^{1/2}+q^{-1/2}) \u\d\u +q^{1/2} \u\u\d \r   \\
           |-q^{1/2} \u\d\d  + (q^{1/2}+q^{-1/2}) \d\u\d -q^{-1/2} \d\d\u \r,
\eer
and another  Doublet II:
\ber
|-q^{-1/2} \d\u\u + (q^{1/2}-q^{-1/2}) \u\d\u +q^{1/2} \u\u\d \r  \\
|q^{1/2} \u\d\d  + (q^{1/2}-q^{-1/2}) \d\u\d -q^{-1/2} \d\d\u \r.
\eer

So, for $q=-1$, all wavefunctions are of mixed symmetry.
In general, for ${\bf 2}^{\otimes n}$, the (spin $n/2$) {\bf n+1}-plet
has for its second-from-the-top level:
\ber
|\sum_{k=1}^{n}  q^{{n+1-2k\over 2}}  \u\u... ...\d... ...\u\u\r \qquad . \\
                                              (k\hbox{th position})~ \nonumber
\eer
Here, always for $q=-1$, the signs of the successive terms alternate,
trailing each permutation with a neighboring isodoublet. This is reminiscent
of fermions, but, again,  aligned isospins do not annihilate,
and hence the representations do not chop up to smaller ones.
{}From the structure of the statistics-altering operator \m
coproduct it is evident that, in this representation, any aligned isospins
are symmetric to each other, while any two anti-aligned ones are antisymmetric
with respect to each other.  The basic structure and dimensionality of
multiplets remains the same as in the classical Lie algebra limit.
This is in contrast to the well-known feature of graded algebra
representations,
whose dimensions differ from those of the corresponding representations
of bosonic Lie algebras (the symmetric representations shrink in size, while
the antisymmetric ones expand).

The features discussed here, by relaxing standard exchange-operator
restrictions, may well be useful in constructing new invariant hamiltonians.
Note, however, that if the states examined are conventional excitations
subject to either Fermi or Bose statistics, naively the multiplets discussed
chop up due to incompatible symmetries: if fermions, then the
extreme members of the triplet are trivialized, just like the singlet; if
bosons, the middle one. Nevertheless, the effect could be reversed by {\em
in addition} introducing a supersymmetric structure on top of the \su.
The reader will no doubt trace this problem to the well-known difficulty in
constructing \su-invariant  field theoretic lagrangians for generic $q$:
Bose or  Fermi symmetry mix (via symmetrization or antisymmetrization) the
singlet $\alpha$ with the triplet, and hence spoil invariance.
Nevertheless, for more involved contexts such as spin-chains or anyonic
systems,
the alternate exchange symmetry structures may allow new constructions,
Boltzmann-weight arrangements, and whence partition functions,
presently under investigation.

\bigskip

{\em We thankfully acknowledge conversations with P. Freund, A. Polychronakos,
J. Uretsky, and S. Vokos.}
\bigskip 

\section*{References}
\begin{enumerate}  \frenchspacing
\item[1.]   V.\ Drinfeld, Sov.Math.Dokl.\ {\bf 32} (1985) 254;
{\it Proc.Int.Cong.\ Mathematicians}, Berkeley 1986, (1987) 798.
M. Jimbo, Lett.Math.Phys.\ {\bf 10} (1985) 63.

\item[2.]  W. Pusz and S. Woronowicz, Rep.Math.Phys.
{\bf 27} (1989) 231. T. Curtright and G. Ghandour, unbublished.
M. Salam and B. Wybourne, J.Phys. {\bf A24} (1991) L317.

\item[3.]  A.\ Polychronakos, Mod.Phys.Lett. {\bf A5} (1990) 2325.

\item[4.] V. Spiridonov, in {\em Quarks '90}, V. Matveev et al. (eds), 1991,
World Scientific, p. 233; Z. Chang, H. Guo,  and H. Yan, Comm.Theor.Phys.
{\bf 14} (1990) 475;
M.\ Chaichian and P.\ Kulish, Phys.Lett. {\bf 234B} (1990) 72;
O. Greenberg, Phys.Rev. {\bf D43} (1991) 4111.

\item[5.] F. Wilczek, Phys.Rev.Lett. {\bf 49} (1982) 957;\\
 Y-H. Chen, F. Wilczek, E. Witten, and B. Halperin, Int.J.Mod.Phys.
{\bf B3} (1989) 1001.\\
C. Aneziris, A. Balachandran, and D. Sen, Int.J.Mod.Phys. {\bf A6} (1991) 4721;
{\em ibid.} {\bf A4} (1989) 5459.

\item[6.] D. Fairlie and C. Zachos, in {\em Quantum Field Theory,
Statistical Mechanics,  Quantum Groups, and Topology}, NATO ARW Series,
T. Curtright et al. (eds.), World Scientific, 1992. Also see C. Zachos in
 {\em Differential Geometric Methods in Theoretical Physics XX}, S. Catto
and A. Rocha (eds.), World Scientific, 1992.

\item[7.] H. Saleur, Nucl.Phys. {\bf B336} (1990) 363. Also see
H. Hinrichsen and V. Rittenberg, preprint CERN-TH.6411/92, February 1992.

\item[8.]  T. Curtright and C. Zachos, Phys.Lett. {\bf 243B} (1990) 237.

\item[9.]  P.\ Roche and D.\ Arnaudon, Lett.Math.Phys.\ {\bf 17} (1989) 295.
For a concise review, see C. Zachos,  ``Paradigms of Quantum Algebras", in
{\em Symmetries in Science V}, B.\ Gruber et al. (eds.), Plenum, 1991, p. 593.

\item[10.] L. Mezincescu and R. Nepomechie, Phys.Lett. {\bf 246B} (1990) 412.

\item[11.]  T. Curtright, G. Ghandour and C. Zachos, J.Math.Phys.
{\bf 32} (1991) 676.

\item[12.] P. Freund and I. Kaplansky, J.Math.Phys {\bf 17 } (1976) 228.

\end{enumerate}
\end{document}